\documentclass[preprint]{aastex}
\usepackage{epsf}
\usepackage{emulateapj5}
\usepackage{onecolfloat}
\usepackage{apjfonts}
\usepackage{amsmath}

\def\gtsim {\lower .1ex\hbox{\rlap{\raise .6ex\hbox{\hskip .3ex
        {\ifmmode{\scriptscriptstyle >}\else
                {$\scriptscriptstyle >$}\fi}}}
        \kern -.4ex{\ifmmode{\scriptscriptstyle \sim}\else
                {$\scriptscriptstyle\sim$}\fi}}}

\newcommand{\be}{\begin{equation}}
\newcommand{\eq}{\end{equation}}

\newcommand{\Msun}{{\rm M_\odot}}

\begin{document}
\submitted{The Astrophysical Journal, submitted}
\vspace{1mm}
\slugcomment{{\em The Astrophysical Journal, submitted}}

\shortauthors{MACCI\`O ET AL.}
\twocolumn[
\lefthead{The origin of polar ring galaxies}
\righthead{Macci\`o et al.}

\title{The origin of polar ring galaxies: evidence for 
galaxy formation by cold accretion}

\author{Andrea V. Macci\`o\altaffilmark{1}, Ben Moore\altaffilmark{1} \& Joachim Stadel\altaffilmark{1}}

\begin{abstract}
Polar ring galaxies are flattened stellar systems with an extended ring of gas
and stars rotating in a plane almost perpendicular to the central
galaxy. We show that their formation can occur naturally in a hierarchical
universe where most low mass galaxies are assembled through the accretion
of cold gas infalling along megaparsec scale filamentary structures.
Within a large cosmological hydrodynamical simulation we find
a system that closely resembles the classic polar ring galaxy NGC 4650A.
How galaxies acquire their gas is a major uncertainty in models of galaxy
formation and recent theoretical work has argued that cold accretion 
plays a major role. This idea is supported by our numerical simulations and
the fact that polar ring galaxies are typically low mass systems.

\end{abstract}

]

\altaffiltext{1}{Institute for Theoretical Physics, University of Z\"urich,
Winterthurerstrasse 190, CH-8057 Z\"urich, Switzerland; andrea@physik.unizh.ch.}

%--------------------------------------------------
\section{INTRODUCTION}
\label{section:introduction}
%-------------------------------------------------

Polar ring galaxies are in general S0 galaxies that are surrounded
by a perpendicular ``polar'' ring or disk of stars and gas
\citep{Schweizer1983,Sparke1986,Whitmore1987,Whitmore1990,Rubin1994}. Less than one percent
of S0's have polar rings, however these features are typically only seen edge on since they
are very low surface brightness, therefore the true fraction of galaxies with
such features is likely to be higher (Figure \ref{fig:hub}(a) shows a classic example).
The currently favored scenario for their formation is a 
merger event between two galaxies, such as the accretion of a large gas
rich satellite which is disrupted in a ring surrounding the central galaxy
\citep{Schweizer1983,Reshetnikov1997,Bekki1998,Tremaine2000,Sparke2000,Bournaud2003}. 
However, the observed mass of baryons in the ring can be comparable to the mass of baryons
in the central disk implying that a major merger must have occurred. 
Similar mass mergers lead to coalescence of the two haloes and central galaxies, 
the remnant violently relaxating into a spheroidal system elongated in the plane of the merger. 
This is now the standard model
for the formation of elliptical galaxies \citep{Toomre1972,Gerhard1981,Barnes1996,BurkertNaab}
therefore the initial conditions must be fine tuned in order to create a polar ring 
galaxy from a merger event.

Galaxy formation is a major unsolved problem that brings together many aspects 
of astrophysics. Observations across many scales and redshifts,
from the cosmic microwave background to the large scale clustering 
of galaxies has defined a standard cosmological model which sets the background
cosmology and initial conditions within which we can try to understand the formation
of cosmic structures (e.g. Tegmark et al. 2004). Within this framework, it is well understood
how the dark matter structures collapse and form galactic haloes,
but much more complex is the behaviour of the baryons. 
Even well posed questions remain unanswered, such as how the gas settles into
the centres of dark matter haloes \citep{Hoyle1953,Dekel2003,Katz03,Keres2005}.

The models that have been proposed for how galaxies get their gas are quite diverse. 
The oldest idea is that the baryons are initially shock heated during an early
merging and virialization phase of the dark matter haloes. The gas begins to 
settle quietly into the central halo in a cooling
flow as the ionised gas in galactic haloes radiates photons, losing energy and gaining
angular momentum. Observational support for hot gaseous haloes around galaxies is lacking 
\citep{Benson2003} but see \citep{tobias05,Mo2005}.
More recently, a wealth of observational evidence suggests that a significant
reservoir of cold gas exists within galactic haloes and groups of galaxies 
(e.g. Westmeier et al. 2005). This cold
material may be stripped from infalling satellites or be infalling directly 
via a smooth cold flow focused along filamentary structures.

%--------------------------------------------------
\section{Numerical Simulations}
\label{section:numsim}
%-------------------------------------------------

Numerical simulations provide the machinery to evolve the initial
conditions through the non-linear regime. We imprint linear fluctuations on the 
dark matter and baryons using a two component particle distribution 
according to expectations from the standard $\Lambda$CDM model.
The simulations were performed with GASOLINE, a multi-stepping, parallel 
TreeSPH $N$-body code (Stadel 2001, Wadsley et al. 2004)  
which can follow the shock heating and radiative cooling processes.
We include radiative and Compton cooling for a primordial 
mixture of hydrogen and helium. 
The gas can form
stars if the physical conditions are appropriate and stellar evolution can feedback energy
into the interstellar and intergalactic medium through supernovae explosions.
The star formation algorithm is based on 
a Jeans instability criteria (Katz 1992), where gas particles in dense, 
unstable regions and in 
convergent flows create star particles at a rate proportional to the local dynamical
time (see also Governato et al 2004). The star formation efficiency
was set to $0.1$, but in the adopted scheme its precise value has only a minor effect 
on the star formation rate (Katz 1992). The code also includes supernova
feedback as described by (Katz 1992), and a UV background following Haardt \&
Madau (1996).

We first perform a dark matter only simulation of a cubic region of a 
concordance  LCDM universe ($\Lambda$=0.7,$\Omega_0$=0.3,
$\sigma_8$=0.9) that is 90 Mpc on a side. From this we 
selected a smaller volume that by a redshift $z=0$ forms a large
candidate Galactic mass halo ($M_{dm} \approx 10^{12} \Msun$) and several
lower mass satellites and surrounding galaxies.  This region is traced back to
the initial conditions and repopulated with many more lower mass 
particles (c.f. Katz \& White 1993), including a gaseous component
throughout the entire high resolution region. The mass per particle of the dark matter
and gaseous particles are respectively
$m_{d} = 4.90 \times 10^5 M_{\odot}$ and $m_g = 9.75 \times 10^4
M_{\odot}$. The dark matter has a spline gravitational softening length of 100 pc and we
have about $4\times 10^6$ particles for each component (dark and gas) in the
high resolution region.

\begin{figure*}[t]
\includegraphics[width=\textwidth]{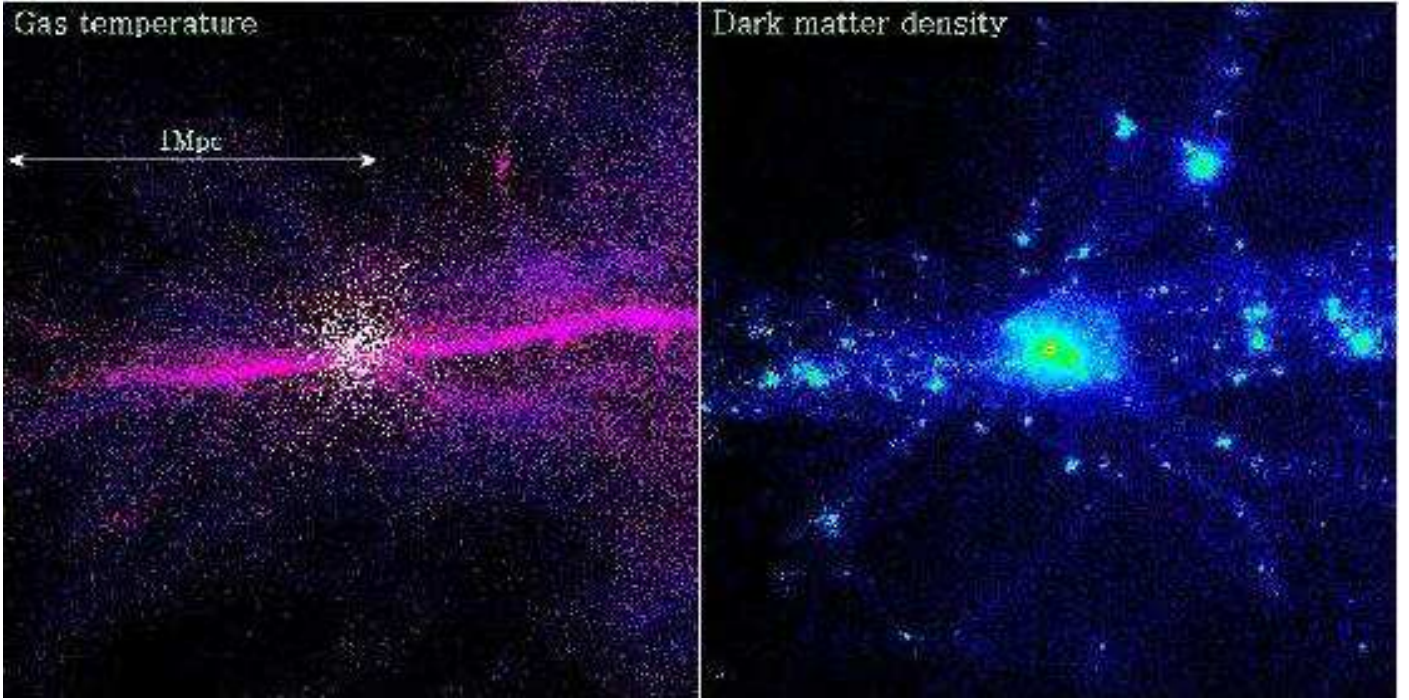}
%\centerline{\epsfxsize=3.2in \epsffile{Image24.ps}}
\caption{The distribution of gas and dark matter in a 2 Mpc wide slice of 
our simulation centered on a galactic mass dark matter halo. The gas temperature is
coloured purple-white in a logarithmic temperature scale from $10^4-10^6K$. 
The narrow filament
of gas flowing into the halo is at a temperature of $\sim2\times10^4K$. The filament in
the dark matter is significantly wider than the gas and
consists of a smooth component and many smaller haloes.
The entire simulation volume is 90 Mpc$^3$ 
and is centred on a region forming several galaxies which are simulated at a higher
resolution than the rest of the cube. The masses and softening of the dark matter
and gas particles are $4.90 \times 10^5 M_{\odot}$, $9.75 \times 10^4
M_{\odot}$ and $0.2$, $0.15$ kpc respectively. The polar ring would appear
face-on in this projection, perpendicular to the filament}
\label{fig:big}
\end{figure*}

%--------------------------------------------------
\section{Results and Discussion}
\label{section:res}
%-------------------------------------------------

Figure \ref{fig:big} shows the final distribution of gas and dark matter within a 
2 Mpc cube surrounding the central halo that hosts our candidate polar ring galaxy.
As pointed out
by previous authors \citep{Dekel2003,Katz03,Weinberg02}, we find
that the dominant mode of galaxy formation is via the accretion of cold gas
along filamentary structures. 
The gas in the filaments is not completely cold, but is heated 
to 15,000K from the collapse and formation of these structures. Some of the
gas is shock heated to the halo virial temperature but most
is accreted in this warm phase.
In our simulation volume with a dozen high resolution galaxies, this is the only system 
with a long lived polar ring, consistent with the observed abundance of these galaxies.

\begin{figure}[t]
\centerline{\epsfxsize=3.2in \epsffile{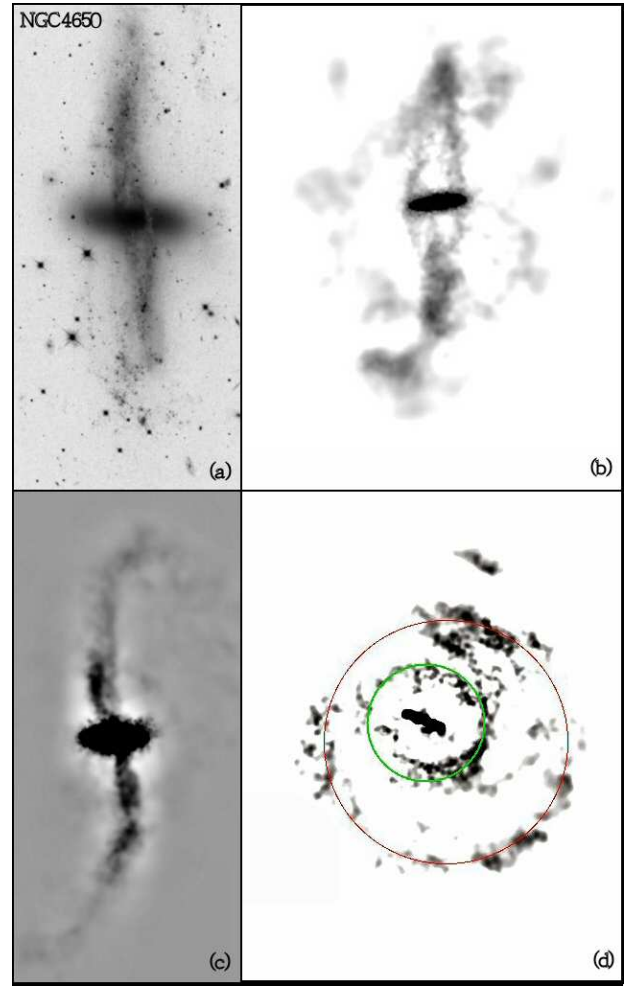}}
\caption{Panel (a) The Hubble Space Telescope image of the classic
polar ring galaxy NGC 4650A (side-length: 24 kpc), courtesy of the Hubble
Heritage site, STSCI. Panels
(b-d) show the smoothed images of the stars and gas of the simulated galaxy in
different projections (side-length 35 kpc): 
(b) inclined 20 degrees (c) the ring is edge on and (d) face on (this Figure
has the same axis orientation of Figure \ref{fig:big}).
The face on image shows a clumpy distribution of gas that lies in two
slightly offset rings.}
\label{fig:hub}
\end{figure}

Figure \ref{fig:hub}(a) shows the hubble space telescope image of NGC 4650A, a classic example
of a polar ring galaxy \citep{Sersic1967}.
Recent observations of its kinematics confirms
that the ring of stars and gas rotates around the central S0 galaxy
\citep{Arnaboldi97,Gallagher2002}. Figure \ref{fig:hub}(b-d) show
close up images of the galaxy that is forming at the centre of the halo in
Figure \ref {fig:big}.
The central galaxy is a rotationally supported
disk that could be classified as an S0 due to its featureless 
``thick disk'' morphology. The galaxy has an extended ring of gas (and stars) 
in near circular motion perpendicular to the disk similar to NGC 4650A.
The face on view of the ring clearly shows that the morphology of the ring is
clumpy and diffuse.

The length of the S0 major axis is 5.1 kpc, and the major/minor axis ration is 
$\approx 9.5$. The S0 baryonic mass is $2.95 \times 10^9
\Msun$, with 17\% of gas.
The ring is mainly made by gas, the total mass (gas+stars, not considering
the S0 galaxy) is $1.23\times 10^9 \Msun$ (inside a cylinder with a radius of
18 kpc and a high of 7 kpc, centered on the S0 galaxy), with a $M_g/M_s$ ratio of about 5.6.
In the same volume the dark matter mass is $6.75 \times 10^9 \Msun$.
The dark matter halo of this galaxy is quite extended, and its virial radius 
(defined according to a spherical overdensity $\Delta=95$) is 98 kpc, 
for a total mass of $5.25 \times 10^{10} \Msun$

%stability
To test test the stability of the ring we cut a box of 9 Mpc side around our
galaxy and then we let it evolve in the future. 
The ring is still present after $\approx 1$ Gyrs, 
then it is destroyed by the dynamical effects of a massive ($1.1 \Msun$)
merging satellite, if we remove all the accreting satellites around our
galaxy the PR survives up to 1.6 Gyrs.
%stability

The ring forms from cold gas that flows along the extended $\sim$1 Mpc filament into
the virialised dark matter halo. The gas streams into the centre of the halo on an
orbit that is slightly offset from purely radial infall and is ``braked'' by ram pressure
drag from the ionised halo gas. 
As it reaches the centre it
impacts with gas in the halo of the existing galaxy and with the
warm gas flowing along the opposite filament. The small impact parameter
is sufficient to leave the gas with sufficient angular momentum that it can
rotate at the halo circular velocity at 10-20 kpc. The ring is clumpy and 
where it has a high enough density it can fragment and form stars according to our
prescription of star formation \citep{katz92}.

\begin{figure}[t]
\centerline{\epsfxsize=3.2in \epsffile{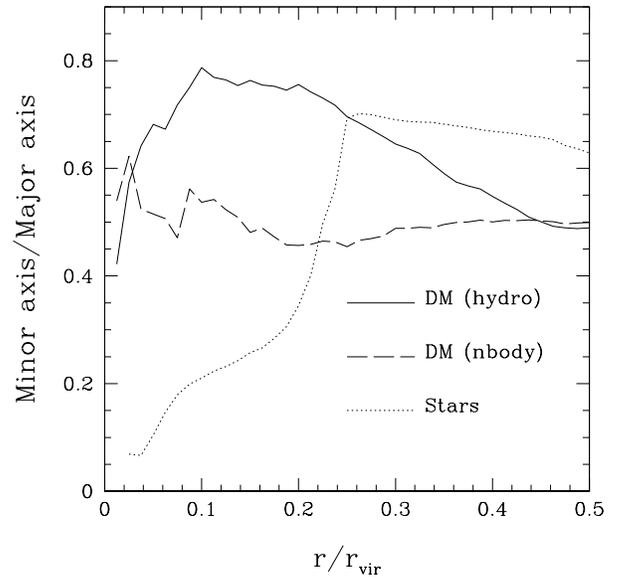}}
\caption{The shape of the dark matter and gaseous components of the 
simulated polar ring galaxy. The dashed curve shows the shape of
the dark matter halo in a simulation where the gas is not allowed to cool.
The solid and dotted curves show the shape of
the dark matter and stellar components in the full hydro-dynamical simulation.
The dark halo becomes rounder but its shape stays quite triaxial in the vicinity
of the disk. The stars are highly flattened indicating the central disk component.}
\label{fig:shape}
\end{figure}

The kinematics of the S0 and the inclined ring can be used to measure the
mass distribution of the stars, gas and dark matter within two planes and 
thus the shape of the potential can also be constrained
\citep{Schweizer1983,Reshetnikov,Arnaboldi94,Sackett94,Combes96,Iodice03}. 
Figure \ref{fig:shape} shows the shape of the dark matter halo and stellar component of the
simulated galaxy. The effect of dissipation alters the shape of the
dark matter distribution \citep{Dubinski1994,Kazantzidis2004}.
At the radius of the ring the halo is closer to spherical,
but the potential (from the baryons and dark matter) is oblate since the stellar
component is rotationally flattened.

The potential at $\sim$20 kpc is still oblate with a flattening $q=0.9$.
Therefore any gas falling away from the perpendicular symmetry axis will rapidly
precess into the plane of the disk \citep{Weinberg02,Muralietal2002}. Only gas accreting
perpendicular to the major axis of the oblate potential will survive for more than a few
dynamical times. Indeed, the ring is long lived in our simulation.
This mode of gas accretion will usually be off-axis and 
provides a source of material from which the disk can grow and form new stars.
Gas accreting onto a bulge or spheroidal would likely produce an S0 galaxy. 
Calculations of larger volumes are required to study the statistics and properties
of polar ring and normal 
galaxies that form via cold accretion. We expect that there is a wide variety
in the properties of such systems, from the mass ratios to their kinematical states. 
Accretion of cold gas could also explain other strange kinematical 
features in galaxies, such as 
counter-rotating stellar and gaseous disks \citep{Merrifield1994,Bertola1996}.

\begin{figure}[t]
\centerline{\epsfxsize=3.2in \epsffile{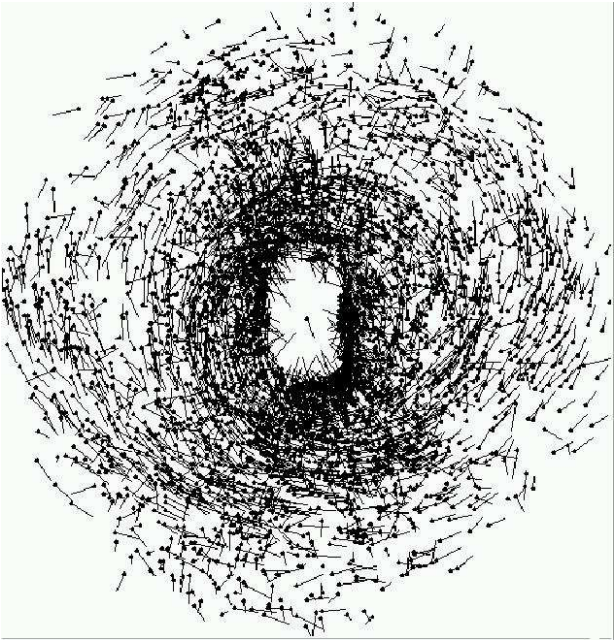}}
\caption{Rotation of the stars in the polar ring. Face-on projection,
   same of figure 2d but rotated of $\approx$ 75 degrees clockwise, 
in order to have the galaxy parallel to the the y-axis; side-length 40 kpc. 
The central region is not plotted}
\end{figure}
\label{fig:rot1}
\begin{figure}[t]
\centerline{\epsfxsize=3.2in \epsffile{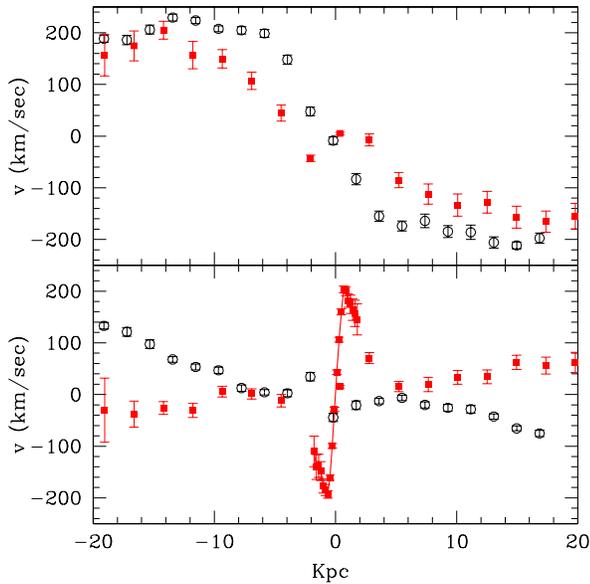}}
\caption{Line of sight velocity of gas (open circles) and stars (solid
  squares). In both cases the slit is along the major axis of the galaxy. 
  In the upper panel the ring is seen edge-on, in the lower face-on. 
  In the lower panel is clearly visible the  spike in the los velocity due to 
  the S0 galaxy.}
\label{fig:rot2}
\end{figure}

\acknowledgments
We thank Linda Sparke for motivating us to investigate
the origin of these fascinating galaxies. We also acknowledge useful
discussions with M. Arnaboldi, J. Bland-Hawthorn, S. Ka\-zant\-zi\-dis and C. Mastropietro.
James Wadsley has carried out most of the development of the
GASOLINE SPH code used and the simulations were carried out on the zBox
supercomputer at the University of Zurich.

\end{document}